# Coexisting Z-type charge and bond order in metallic NaRu$_2$O$_4$

Arvind Kumar Yogi[1,2,3*,#], Alexander Yaresko[4*], C. I. Sathish[1,2], Hasung Sim[1,2], Daisuke Morikawa[5], J. Nuss,[4] Kenji Tsuda[6], Y. Noda[5,7], Daniel I. Khomskii[8], and Je-Geun Park[1,2,9,10#]

[1]*Center for Correlated Electron Systems, Institute for Basic Science (IBS), Seoul 08826, Korea*
[2]*Department of Physics and Astronomy, Seoul National University, Seoul 08826, Korea*
[3]*UGC-DAE Consortium for Scientific Research, Indore-452001, India*
[4]*Max-Planck-Institut für Festkörperforschung, 70569 Stuttgart, Germany*
[5]*Institute of Multidisciplinary Research for Advanced Materials, Tohoku University, Sendai 980-8577, Japan*
[6]*Frontier Research Institute for Interdisciplinary Sciences, Tohoku University, Sendai 980-8577, Japan*
[7]*J-PARC Center, Japan Atomic Energy Agency, 2-4 Shirakata, Tokai, Ibaraki, 319-1195, Japan*
[8]*Institute of Physics II, University of Cologne, 50937 Cologne, Germany*
[9]*Center for Quantum Materials, Seoul National University, Seoul 08826, Korea*
[10]*Institute of Applied Physics, Seoul National University, Seoul 08826, Korea*

\* Authors with equal contributions
\# Corresponding authors: jgpark10@snu.ac.kr & yogi.arvind2003@gmail.com

How particular bonds form in quantum materials has been a long-standing puzzle. Two key concepts dealing with charge degrees of freedom are dimerization (forming metal-metal bonds) and charge ordering (CO) [1,2]. Since the 1930s, these two concepts have been frequently invoked to explain numerous exciting quantum materials [3-6], typically insulators. Here we report dimerization and CO within the dimers coexisting in metallic NaRu$_2$O$_4$. By combining high-resolution x-ray diffraction studies and theoretical calculations, we demonstrate that this unique phenomenon occurs through a new type of bonding, which we call Z-type ordering. The low-temperature superstructure has strong dimerization in legs of zigzag ladders, with short dimers in legs connected by short zigzag bonds, forming Z-shape clusters: simultaneously, site-centered charge ordering also appears. Our results demonstrate the yet unknown flexibility of quantum materials with the intricate interplay among orbital, charge, and lattice degrees of freedom.

Chemical bonding lies at the heart of all natural sciences like condensed matter physics, chemistry, and even biology. As such, the understanding of chemical bonding was a great puzzle for a long time. In the long and rich history of condensed matter physics, one cannot emphasize strongly enough the importance of the two basic concepts: dimerization [1] and charge ordering (CO) [2], both born in the 1930s. However, these two phenomena have been rarely found to coexist in one compound, much less in metallic systems.

Many studies in the late 1990s and early 2000s demonstrated, one example after another, how the new patterns emerge out of simple structures: hexamers in $ZnCr_2O_4$ [3], octamers in $CuIr_2S_4$ [4], spontaneous dimensionality reduction in $Tl_2Ru_2O_7$ [5], and the formation of trimeron in $Fe_3O_4$ [6], to name only a few. Despite the various properties, an ever occurring theme is that these phenomena typically happen near the metal-insulator boundary, with the ordered states usually insulating. This characteristic feature is naturally captured by Mott-Hubbard physics [7], and it is particularly prominent among frustrated magnets [8-11]. It is relatively less well-known how this rich physics plays out in metallic systems; it remains unchartered territory to date. One can imagine that this question may become more important in the newly emerging field of quantum materials that increasingly include metallic systems [12].

Ru has a special place in this extensive search because of its two characteristic features: the modest spin-orbit coupling, not too small nor too large compared to the Coulomb energy and the bandwidth; another is a relatively stronger importance of orbital physics [7]. This combination of the two has been behind the active researches on Ru oxides; for instance, superconducting $Sr_2RuO_4$ and ferromagnetic $SrRuO_3$ are two prominent examples [13,14]. In comparison, there have been few studies on Na-Ru-O systems, including $NaRu_2O_4$. Using neutron diffraction, the authors of Ref. [15] determined that its room-temperature structure has the space group of *Pnma*, the typical orthorhombic structure among oxides: to our best knowledge, it was initially reported in 1975 [16] before being re-examined in 2006. It was reported to have a paramagnetic susceptibility down to the lowest temperature. While studying another Na compound, $Na_{2.7}Ru_4O_9$ [17] which was also investigated in the same Ref. [15], we realized that the high-temperature phase of these Na-Ru-O systems could exhibit exciting new unreported properties and hence deserve careful studies.

In this report, we made extensive studies of both structure and physical properties of $NaRu_2O_4$ over a wide temperature range up to 600 K above which it becomes chemically unstable. We made the high-resolution XRD study above room temperature to find a surprising first-order phase transition during this study. We found that $NaRu_2O_4$ undergoes a metal-metal phase transition below 535 K with a distinct structure change, from a $CaFe_2O_4$-type structure with the orthorhombic *Pnma* symmetry at a higher temperature to a monoclinic structure *P112$_1$/a* below $T_c$. By combining experimental and theoretical studies, we found that at this transition we have simultaneously a site-centered charge ordering coexisting with dimerization.

## Results

**Bulk properties.** Electrical resistivity (ρ) measurements were carried out using a homemade system equipped with a furnace (300 to 685 K) and a pulsed-tube cryostat with the base temperature of 3 K. The electrical resistance was measured in a four-point geometry, where contacts were made using silver paint and 25 μm gold wire as shown in the inset of Fig. 1a. The current was applied perpendicular to the single crystal length, which is the crystallographic *b*-axis. As the single crystals are all in a think needle form, we could not measure the resistivity along the other crystallographic directions. We tested several long rod-shaped as-grown crystals and discovered the same transition (see also Fig. S1). The resistivity data shows that a phase transition occurs in NaRu$_2$O$_4$ at $T_c$ = 535 K with small hysteresis in the electrical resistivity, a sign of a weak first-order phase transition at $T_c$. We will extensively discuss this transition below; it is the main feature of NaRu$_2$O$_4$.

The resistivity $\rho$ (T) data measured down to 3.3 K indicates metallic behavior with a room temperature resistivity of ~ 40 μΩ-cm. To have a quantitative understanding, we analyzed the results theoretically using the Bloch-Grüneisen (BG) model. The following formula gives the expression for the temperature-dependent part of the metallic resistivity:

$$\rho \approx \left(\frac{3}{\hbar e^2 v_F^2}\right) \frac{k_B T}{M v_S^2} \int |v(q)|^2 \left[\frac{(\hbar\omega/k_B T)^2 q^3 \mathrm{d}q}{(\exp(\hbar\omega/k_B T)-1)(1-\exp(-\hbar\omega/k_B T))}\right] \quad (1),$$

where *v(q)* is the Fourier transform of the potential associated with one lattice site and $v_s$ being the sound velocity. When Eq. (1) was considered for the acoustic phonon contribution, it leads to the BG function as shown below:

$$\rho_{e-\mathrm{ph}}(T, \theta_D) = 4A_{\mathrm{ac}}(T/\theta_D)^4 \times T \int_0^{\theta_D/T} x^5 (e^x - 1)^{-1}(1 - e^{-x})^{-1} \mathrm{d}x \quad (2),$$

where $x = \hbar\omega/k_B T$, $\theta_D$ is Debye temperature, and $A_{ac}$ is a proportionality constant. The total resistivity of a material can be written as,

$$\rho(T) = \rho_0 + \rho_{e-\mathrm{ph}}(T, \theta_D),$$

where $\rho_0$ is the temperature-independent residual resistivity and the second term represents the electron-phonon scattering. The B-G model is found to work well for our study, and the theoretically estimated $\rho$ is quite consistent with the experimental data for NaRu$_2$O$_4$. The characteristic Debye temperature ($\theta_D$) and residual resistivity ($\rho_0$) is found to be 452.2 K and 35.9 μΩ-cm, respectively.

We also measured magnetic susceptibility and specific heat of NaRu$_2$O$_4$ in the accessible temperature range of 2 to 300 K. The susceptibility ($\chi$) for NaRu$_2$O$_4$ was measured in the temperature range between 1.9 ≤ *T* ≤ 300 K with an 0.5 T of the external field applied perpendicular to the length of the as-grown crystal as in Fig. 1a. No magnetic ordering was observed in the measured temperature range. It is noticeable that the susceptibility is almost temperature-independent over the wide temperature range, showing a sizeable paramagnetic behavior consistent with the reported susceptibility data on the polycrystalline sample [15]. The low-temperature upturn is apparently due to some magnetic impurities, and it can be explained by using a modified Curie-Weiss formula (see the red line in Fig. 1b and SI Note II). The fit yields the following values: Curie-Weiss temperature $\theta_{CW}$ = -1.8 K, observed moment $\mu_{eff}$ = 0.04 $\mu_B$ per Ru atoms and $\chi_0$ = 1.15×10$^{-7}$ (cm$^3$/ Ru mol)$^{-1}$, where the latter is the temperature-independent van Vleck contribution to the susceptibility. The CW-fit gives a significantly less observed moment, indicating negligible localized magnetic moments in NaRu$_2$O$_4$, similar to the reported polycrystalline sample [15].

The heat capacity was measured between 1.9 $\leq T \leq$ 300 K and is displayed in Fig. 1c. The heat capacity did not show any $\lambda$-type anomaly over the entire temperature range, confirming no long-range magnetic order from 300 to 1.9 K. Low-temperature heat capacity data were analyzed using the usual formula of $C_p = \gamma T + \beta T^3$, where $\gamma$ is the electronic specific-heat coefficient and $\beta$ is the phonon contribution at low temperature (see the inset of Fig. 1c). The electronic contribution to the specific heat ($\gamma$) for NaRu$_2$O$_4$ was 3.93 mJ/mol K$^2$, consistent with the metallic behavior.

**Single-crystal XRD data & crystal structures at high and low temperatures.** To understand the high-temperature phase transition seen in the resistivity data, we conducted high-resolution single-crystal diffraction experiments to obtain the most informative data. We performed the temperature-dependent SC-XRD experiment from 300 to 575 K using a single crystal diffractometer (XtaLAB P200, Rigaku) (see Fig. 2). In our room-temperature data, we observed **q** = (0, ½, 0) superlattice peak (satellite reflections). It is important to note that this presence of the satellite reflection cannot possibly be explained by the *Pnma* space group proposed in Ref. [15].

While searching battling for a possible true structure at room temperature, we decided to employ the CBED (Convergence Beam Electron Diffraction) as it is the most powerful technique in determining the space group and gives crucial information about the lattice's symmetry [18] (see Fig. 3). The weak superlattice reflections in the x-ray diffraction experiments already form strong evidence for broken symmetry (both translation and rotational) and lattice distortion in a crystal. In addition, the superlattice peak at **q** = (0, ½, 0) disappears upon heating just above $T_c$. We collected CBED diffraction patterns along the various [100], [130], [001], and [201] axes as shown in Fig. 3(a-d) at room temperature. The CBED experiment revealed *P112$_1$/a* symmetry (no. 14 with setting unique axis *c* [19]) at room temperature, a subgroup of the reported *Pnma* symmetry. Therefore, a modified structure model under group-subgroup relation is used to solve the room temperature SC-XRD data. We used the Shelx software [20] to address the twin problem in our refinement. In the twin refinement for NaRu$_2$O$_4$ SC-XRD data, we used a monoclinic merohedral twin model under twin law as (-100) (010) (00-1) to get better refinement parameters and statistics as well (see SI Note III). All the reflection patterns collected at 300 K could be indexed as monoclinic *P112$_1$/a* symmetry with a primitive lattice of *a* = 9.273 (6) Å, *b* = 5.643 (3) Å, and *c* = 11.17 (7) Å.

Table-1 summarizes the SC-XRD refinement results for the LT-phase (300 K). All the crystallographic sites were considered to be fully occupied and were kept fixed during the refinement. We carried out the structure analysis by including the twin structure of monoclinic cells. The structural parameters were finally determined from the least-squares refinement of the single-crystal XRD data. For the single-crystal refinement, we used a total of 5021 reflections, and all the reflections were well indexed by a monoclinic $a_0 \times 2b_0 \times c_0$ cell. We used *P112$_1$/a* (no. 14) with a table setting choice: *c1*. No constraints were placed on the atomic positions. The comparison of calculated and observed intensity (F$^2$) after the structural refinement at room temperature is shown in Fig. 2a.

As regards the HT phase, our single-crystal data could not uniquely solve the structure above $T_c$ (HT-phase) due to the sizeable diffuse scattering. Most probably, at such high temperatures, lighter Na ions located in 1D tunnels of $NaRu_2O_4$ start to move within the tunnels. In this situation, we decided to begin with positional parameters of distorted phase (Table-2, LT-phase with $a_0 \times 2b_0 \times c_0$ $P112_1/a$ from Shelxl refinement with twin), and generate undistorted positional parameters (undistorted $P112_1/a$ $a_0 \times b_0 \times c_0$). The undistorted positional parameters of $P112_1/a$ symmetry are used to get positional parameters for orthorhombic HT-phase $Pnma$ symmetry under cell-symmetry and group-subgroup relation [19]. Expected coordinates at HT $Pnma$ phase based on LT $P112_1/a$ structure ($a_0 \times b_0 \times c_0$) are shown in Table-2. Excellent relation is seen between expected positions and reported one [15] (by missing superlattice reflections: $a_0 \times b_0 \times c_0$), as shown in Table-3.

The crystal structure of $NaRu_2O_4$ is close to that of the prototype compound $CaFe_2O_4$ with similar coordination environments. In our revised crystal structure for $NaRu_2O_4$, the most characteristic feature we found is the edge shared octahedral double $RuO_6$ chain, which runs along the crystallographic *b*-axis. These double chains form a two-leg zigzag ladders along the crystallographic *b*-axis. The $RuO_6$ octahedra are edge-shared within each chain, tied to the neighboring chain through the corner oxygen. This leads to an interesting lattice geometry with pseudo triangular tunnels, in which a single Na atom can be accommodated, as shown in Fig. 4(a), which is also found in various $CaFe_2O_4$-type lattices [21-26]. Our detailed single-crystal x-ray diffraction analysis shows a transition from the HT (high-temperature) $Pnma$ structure to the LT (low-temperature) $P112_1/a$ structure at $T_c$ = 535 K, as shown in Fig. 4(b, c): which is consistent with the results of resistivity measurements shown in Fig.1a.

According to our diffraction experiments done between $300 \leq T \leq 575$ K, the **q** = (0, ½, 0) superlattice peak (satellite reflection) persist right up to 550 K as shown in Fig. 5a. The line cut from the temperature-dependent reciprocal image analysis for peak **q** = (0, ½, 0) shows clearly the suppression of intensity as a function of temperature as in the main article Fig. 5(b,c). The **q** = (0, ½, 0) superlattice peak is substantially suppressed just above the first-order transition ($T_c$ = 535 K).

**Analysis and discussion**

The structure of $NaRu_2O_4$ can be visualized as the double chain of edge-sharing $RuO_6$ octahedra running in the *b*-direction and forming 2-leg zigzag ladders (see Fig. 6a). The corner-sharing oxygen atoms connect these ladders, also having a zigzag pattern. In effect, there appear in this system, not one, but *three* different types of zigzag ladders: those in edge-sharing double chains (ladders-2 marked by red in Fig. 6b. For shortness, we will call them "red" ladders), and two types of corner-sharing zigzag ladders, marked in Fig. 6(b,c) by blue and green. When describing this class of materials, one often pays primary attention to the phenomena occurring in the "red" ladders (see [27]). But the corner-sharing ladders are sometimes found to play a more critical role in systems like hollandite $K_2Cr_8O_{16}$ [28]. As we will see below, these edge-sharing ladders also play a crucial role in $NaRu_2O_4$.

Here it is worth while to think about possibility of certain effect due to off-stoichiometry. In particular, one can ask whether the observed metallic behaviour is related to a sample quality issue.

For instance, it is known that such stoichiometry plays an important role in determining the low-temperature properties of RuP [29], whose high-quality powder samples exhibit a metal-insulator transition. On the other hand, single crystal of RuP, presumably off-stoichiometric, has a metallic ground state. For NaRu$_2$O$_4$, we can rule out this issue as our powder sample also shows a metallic behaviour (see also Fig. S2). Another point is that our heat capacity measurements taken on two different samples consistently show a finite gamma value at low temperature, which is consistent with the data reported in [15].

An exciting picture now emerges from the detailed structural study. The HT phase has two independent Ru sites (Ru$_1$ and Ru$_2$), equally spaced (2.82 Å) with the lattice constant $b_0$ along two Ru chains. The most prominent feature of the structural changes is the appearance of strong dimerization in Ru chains (legs of zigzag ladders), leading to cell-doubling in the b-direction (see Fig. 7a). Ru-Ru distance in short Ru dimers is 2.60 Å– even shorter than the Ru-Ru distance in Ru metal (2.65 Å). It is a clear sign of direct metal-metal bonding in such dimers – which quite unexpectedly coexist with metallic conductivity. Simultaneously with the dimerization in 1D Ru chains – legs of zigzag ladders, Ru sites themselves become inequivalent, Ru(A) and Ru(B), a clear sign of a site-centered CO in NaRu$_2$O$_4$. The coexistence of bond- and site-centered orderings is extremely rare in metallic systems, making NaRu$_2$O$_4$ especially interesting.

Another essential feature of this superstructure is the difference of distances in the inter-chain bond length (interdimer), both in edge-sharing "red" and in corner-sharing "green" and "blue" ladders (Fig. 6b). The Ru$_1$(A)-Ru$_1$(A) and Ru$_2$(A)-Ru$_2$(A) distance in the edge-sharing ladder is 3.10 and 3.09 Å, while the Ru$_1$(A)-Ru$_1$(B) and Ru$_2$(A)-Ru$_2$(B) distance are 3.12 and 3.11 Å, respectively, with the remaining distance between dimers (Ru$_1$(B)-Ru$_1$(B) and Ru$_2$(B)-Ru$_2$(B)) being 3.16 Å (see Figs. 7(a,b)). From the viewpoint of the bond order, short bonds have a shape like a "letter-Z" (see Fig. 8). The same is true for two other types of ladders, edge-sharing "blue" and "green" ladders. We thus see that the dimers also form Z-type orbital clusters in corner-sharing ladders. The relative difference of interdimer bonds in these is even more prominent than in the "red" edge-sharing ladders: it is ~ 1.7% compared to ~ 0.7% in the "red" ladders. This already hints that what happens in corner-sharing ladders is probably more critical than in edge-sharing ones, as in hollandite K$_2$Cr$_8$O$_{16}$ [28]. According to our ab-initio calculations based on the charge valence of our choice [Ru$_1$(A)$^{3+}$, Ru$_1$(B)$^{4+}$, Ru$_2$(B)$^{4+}$, Ru$_2$(A)$^{3+}$], the nn hoppings in corner-sharing blue and green zigzags are: for blue corner sharing bonds t=0.40 eV, for green corner sharing ones t=0.29 eV, and for red edge sharing bonds t=0.35 eV. For other CO the values are somewhat different but the blue corner sharing bonds are always the strongest. It shows the dominance of the blue bond among all relevant bonds.

For the determination of the oxidation states of cations (Na and Ru), we carried out calculations by using the bond valence sum (BVS) method with an inbuilt program in Fullprof software [30]. The principle of BVS method in determining the valence $V$ from different cation sites [31,32] can be given by an expression $V = \Sigma_i [\exp(d_0-d_i)/0.37]$. Here, $d_i$ is the experimental bond lengths to the surrounding ions, and $d_0$ is a tabulated empirical value characteristic for the cation-anion pair [31,32]. The BVS result for NaRu$_2$O$_4$ was determined for the LT phase. The charge at different Ru sites is illustrated by different colors (brown-Ru$^{4+}$ and green-Ru$^{3+}$) in the main article Figs. 6 and 7. The BVS's for the eight oxygen sites were between 1.86 and 2.20. The BVS's for the four Ru cations sites for different charge configurations (we call them four different models) are shown below in Table-4. The respective reliability parameter as Global Instability Index (GII) is also listed for each BVS model for $NaRu_2O_4$.

To be more technical, an important question is how different Ru valence states are distributed in the low-temperature phase. Specifically, one has to know which ions, in which charge state, are connected by these shorter zigzag rungs in orbital Z clusters. For that, we used the results from the bond valence sum (BVS) analysis at each Ru, an empirical measure of its charge or its valence state [33, 34]. In the HT phase, all Ru are equivalent, and BVS gives the valence $Ru^{3.5+}$. In contrast, Ru ions in a short dimer are indeed inequivalent in the LT phase, with the valence estimated from BVS being 3.536 and 3.106; that is, one can think of $Ru^{4+}$ and $Ru^{3+}$ in each dimer. However, from our structural data alone, although collected with high-resolution single-crystal x-ray diffraction and the total number of 5021 Bragg peaks, we cannot uniquely discriminate the four possible CO patterns, mainly because an x-ray experiment fundamentally suffers from the weak scattering signals of oxygen. The reliability parameters (Global Instability Index (GII)) for four different CO patterns are comparable to one another, about 10-11% (see Table 4). Still, a slightly better index is obtained for charge distribution with $Ru^{4+}$ ions at the ends of short diagonal bonds in blue and green ladders (reliability factor 10.43 % for model 3 in Table 4). In contrast, these are $Ru^{3+}$ in edge-sharing red ladders. Note that different ladders are connected in the structure of $NaRu_2O_4$ so that the charges at short zigzags in neighboring ladders are opposite (see Fig. 8).

**Theoretical Studies**

As we commented above, there is unavoidable ambiguity about our determination of the charge valence using the BVS method although we used the total number of 5021 Bragg peaks from the high-resolution single-crystal x-ray diffraction experiment. Therefore we employed *ab-initio* LSDA and LSDA+*U* calculations to clarify the charge pattern favored by the LT crystal structure. We used different Coulomb repulsion *U* values in the range 2.7 - 4.2 eV and Hund's coupling $J_H$ = 0.7 eV as estimated from LSDA. All calculations were performed for the experimental HT and LT crystal structures. No attempts to optimize lattice geometry have been made. The effect of spin-orbit coupling (SOC) on Ru $t_{2g}$ bands was relatively weak. In the following, for simplicity, we discuss the results of scalar-relativistic calculations. Our theoretical studies show that different self-consistent spin and charge-ordered solutions can be stabilized for sufficiently large *U* (3.7 eV). These solutions have the following common features: Ru $d_{xy}$ states (in local coordinates, with the axes directed from Ru to O) are strongly split into occupied bonding and unoccupied antibonding molecular orbital states (see Fig. 7c-f). Second, each dimer's Ru ions become formally $Ru^{3+}$, with its $d_{xz}$, $d_{yz}$ orbitals being doubly occupied. Another Ru ion of the dimer, $Ru^{4+}$, has one half-occupied orbital, approximated by a linear combination of $d_{xz}$ and $d_{yz}$ lying in the plane perpendicular to the *b*-axis. The $Ru^{4+}$ ion acquires a spin moment of about 0.8 $\mu_B$. Orbital-resolved density of Ru *d* states illustrates the charge disproportionation. With decreasing *U*, the hybridization becomes more robust between the unoccupied $d_{xz}$, $d_{yz}$ orbital of $Ru^{4+}$ and occupied $t_{2g}$ orbitals of $Ru^{3+}$, with a gap separating the unoccupied $d_{xz}$, $d_{yz}$ band from the top of the valence band closes, and for *U* ~ 2.7 eV the charge disproportionation between $Ru^{3+}$ and $Ru^{4+}$ ions disappears (see Fig. 7c-f). It should be noted that this is only a rough estimate for the critical *U* value which may change if optimized crystal structure were used in the calculations and is in general implementation specific.

Out of the four CO patterns allowed for the LT structure, three are nearly degenerate, with the energy difference being less than 3 meV/f.u. This finding agrees with the BVS analysis results, which give comparable reliability factors for different CO. Still, the lowest total energy is found for one CO pattern: for which the lowest value of GII was obtained experimentally from BVS. The final charge pattern is that the short zigzag bonds in Z-cluster in corner-sharing "blue" ladders connect $Ru^{4+}$ - $Ru^{4+}$ ions (see Fig. 8).

Finally, we would like to note that the structure of $NaRu_2O_4$ is composed of frustrated 2-leg zigzag ladders with a spin singlet ground state, which makes this material quite interesting (see SI Note IV). As shown in Fig. 9(a,b), the orthorhombic *Pnma* crystal structure of $NaRu_2O_4$ has two ruthenium atoms $Ru_1$ and $Ru_2$. The edge-shared bond between the two $RuO_6$ octahedra for each dimer 1 (2.60 Å), dimer 2 (2.61 Å) and Z diagonal-bond (3.10 Å) give rise to two Ru–O–Ru bonds which are deviated from 90˚ for each and also all of the edge-shared octahedral shows significant trigonal-distortion. The schematic representation shows the double chain-forming two-leg zigzag ladders (solid thick light-green, reddish-brown, and sky-blue lines) running along the crystallographic b-axis in crystal. The 2-leg zigzag ladders are connected in two different forms: the corner (blue or green ladders) or the edge (red ladders) shared $RuO_6$ octahedra. The $t_{2g}$ energy-level sketch with respective electron fillings is shown for a charge-ordered state, so-called Z-order (left panel) and Ru-dimer (right panel) in Fig. 9d.

**Conclusions**

Thus the following final picture emerges from both the experimental and theoretical studies. $NaRu_2O_4$ has very strong dimerization below $T_c$ = 535 K (in other words, strong metal-metal bonding with the use of *xy*-orbitals) in one-dimensional chains and legs of zigzag ladders (this is a bond-centered CDW). With the average valence $Ru^{3.5+}$, one active electron (or one hole) remains per a dimer. And simultaneously with this dimerization (actually the orbitally-driven Peierls dimerization [35-38]), these extra electrons order in each dimer– there appears also a site-centered charge ordering or site-centered CDW. The remaining electrons form one stronger diagonal bond between dimers in neighboring legs, creating Z-clusters (Fig. 8). The detailed form of orbitals in forming this short zigzag bond in the "blue" corner-sharing ladders is shown in Fig. 9c. But this extra bonding is not strong enough to make the system insulating, and it remains metallic below $T_c$ due to the multi-orbital effect. These Ru-Ru dimers probably survive above $T_c$, forming dimer liquid, e.g., in $Li_2RuO_3$ [39]; this dimer liquid phase in $Li_2RuO_3$ was found stable upon doping too [40]. This could be checked by probes like EXAFS or PDF (Pair Distribution Function) analysis.

To put our results in a broader perspective, we would like to comment on why the dimerization and the site-centered CO coexistence are favorable for metallic $NaRu_2O_4$, while it has been rarely seen in other transition metal compounds: another example is $IrTe_2$ [41]. Most importantly, the Ru 4*d* bands of $NaRu_2O_4$ exhibit the delicate balance between the correlations, the Coulomb *U*, and the bandwidth *W*. Two other factors play a crucial role: orbital freedom, leading to orbitally-driven dimerization, and the mixed-valence of Ru, $Ru^{3.5+}$, promoting site-centered charge ordering. These features seem to be the key to realizing CO and dimer coexistence in a metallic $NaRu_2O_4$. Going further, we anticipate that our observations will prompt renewed efforts towards the hitherto scarcely investigated Mott-Hubbard physics in a metallic regime.

**Acknowledgments**

We would like to thank Juan Rodríguez-Carvajal, Sang-Wook Cheong, and Masahiko Isobe for fruitful discussions. Work at the Center for Quantum Materials was supported by the Leading Researcher Program of the National Research Foundation of Korea (Grant No. 2020R1A3B2079375) with partial funding by the Institute for Basic Science, Republic of Korea. The work of D. Kh. was supported by the Deutsche Forschungsgemeinschaft (DFG, German Research Foundation) - Project number 277146847 - CRC 1238). A.K.Y. acknowledges financial support from the Institute for Basic Science of the Republic of Korea and was partially supported by the Research Fellowships of the Max-Planck Institute Foundation, Germany.

**Authors Contributions**

JGP initiated and supervised the project. AKY, HS, and CIS prepared the samples and carried out the bulk measurements. AKY did the structural analysis under the supervision of YN. DM and KT did the CBED measurements. JN did the merohedral-twin refinement by using Shelx software. AY made the theoretical studies of LDA calculations. DK and JGP interpreted the data with the help of AKY, AY, and YN. AKY, AY, YN, DK, and JGP wrote the manuscript with inputs from all authors.

**Competing financial interests**

The authors declare no competing financial interests.

**Methods**

Sample preparations: Polycrystalline NaRu$_2$O$_4$ samples were synthesized by solid-state reaction of preheated RuO$_2$ (99.999%, Aldrich) and Na$_2$CO$_3$ (99.999%, Aldrich) under an Ar-gas environment at 950 ˚C for 90 h with several intermediate grindings and pelletization. Subsequently, high-quality single crystals of NaRu$_2$O$_4$ were grown from this polycrystalline powder via a modified self-flux vapor transport reaction under flowing Ar-gas (ultra-pure 99.999%). Long needle-shaped high-quality single crystals (1 × 0.1 × 0.1 mm$^3$) were obtained from the final products. The synthesized polycrystalline sample's phase purity was checked using a Bruker D8 Discover diffractometer with a Cu-Kα source with no impurity peaks observed. Elemental analysis was subsequently done confirming the samples' stoichiometry: we used a COXI EM-30 scanning electron microscope equipped with a Bruker QUANTAX 70 energy dispersive x-ray system.

Bulk properties: Electrical resistivity (ρ) measurements were carried out using a homemade system equipped with a furnace (300 to 685 K) and a pulsed-tube cryostat (down to 3 K, Oxford). The electrical resistance was measured in the four-point geometry on a static sample holder, where the contacts to the sample were made using silver paint and 25 μm gold wire. The current was applied perpendicular to the single crystal length, the crystallographic *b*-axis. Magnetic susceptibility χ(T) measurements were taken using an MPMS-SQUID magnetometer (Quantum Design). Heat capacity C$_p$(T) measurements were made using the commercial Physical Property Measurement System (PPMS, Quantum Design).

Structure analysis: The temperature-dependent single-crystal x-ray diffraction (SC-XRD) was performed from 300 to 575 K using a single crystal diffractometer (XtaLAB P200, Rigaku). The room-temperature SC-XRD data exhibit superstructures **q** = (0, ½, 0). The crystal structure was refined by using Fullprof suite software and the *ShelX* program with a modified monoclinic merohedral twin model under twin law as (-100) (010) (00-1) (see SI Note III). We also carried out temperature-dependent SC-XRD measurements on NaRu$_2$O$_4$ crystal to confirm the structural phase transition. Besides, we made a convergent beam electron diffraction (CBED) experiment, which is the most accurate in determining the correct symmetry of materials, especially regarding the loss of the inversion center. This CBED measurement allows us to determine the exact space-group of NaRu$_2$O$_4$. The CBED result for NaRu$_2$O$_4$ gives *P112$_1$/a* symmetry and works well for the observed commensurate superstructures with **q** = (0, ½, 0).

LDA band calculations: Band structure calculations were performed for the experimentally obtained high- and low-temperature crystal structures using the LMTO method as implemented in the PY LMTO computer code [42] and the Perdew-Wang [43] parameterization of the exchange-correlation potential in the local spin density approximation (LSDA). To account for Coulomb repulsion within Ru d shell, we used a rotationally invariant formulation of the LSDA+*U* method [44] with the double-counting term in the fully localized limit. An LDA calculation for the HT crystal structure gave a nonmagnetic metallic solution with partially occupied Ru$_1$ and Ru$_2$ *t$_{2g}$* orbitals. The band structure's characteristic feature is Ru *t$_{2g}$* bands with stable quasi-one-dimensional dispersion along the b direction. These bands originate from Ru *d$_{xy}$* orbitals and are evidence of strong direct Ru *d$_{xy}$*-*d$_{xy}$* hopping along the b chains. Here and in the following, Ru orbital is defined in local frames with the axes directed approximately to three nearest O ions in such a way that the local [110] direction always points along the crystallographic *b*-axis. The LDA band structure calculated for the LT structure is qualitatively different. Although it remains nonmagnetic and metallic, the shortening of the Ru-Ru distance within dimers leads to the splitting of Ru $d_{xy}$ drives bands into completely filled bonding states 2 eV below $E_F$ and unoccupied antibonding ones 1 eV above $E_F$. The bonding $d_{xy}$ bands accommodate one electron per Ru while the remaining 3.5 $d$ electrons fill bands are formed by Ru $d_{xz}$, $d_{yz}$ states.

**Figure Captions**

**Figure 1: Bulk properties. (a)** The electrical resistivity data as a function of temperature is taken over single-crystal with the solid red line for the BG model. Inset shows the single crystal used for the four-probe resistivity method. **(b)** Magnetic susceptibility ($\chi$) is measured at an applied magnetic field $B$ = 0.5 T. The inset shows inverse susceptibility with the modified Curie-Weiss (CW) (the solid red line). **(c)** Temperature dependence of the specific heat $C_p$ measured at zero fields ($B$ = 0 T). Solid black circles are the raw data with $C_p/T$ shown in the inset of the figure and the solid red line for the corresponding $C_p/T = \gamma + \beta T^2$ fitting.

**Figure 2: Room-temperature SC-XRD refinement for LT-phase: Single-crystal XRD reciprocal-lattice images and lattice parameters across the structural transition**. **(a)** The SC-XRD refinement includes a modified structural model with a new space group ($P112_1/a$) determined from CBED experiments. The refinement is performed by including a monoclinic merohedral twin model under twin law as (-100) (010) (00-1). The solid red and blue circles are fundamental and superlattice reflections, respectively. The inset of the figure shows the optical image of the measured crystal. **(b)** Single-crystal x-ray diffraction data were measured at various temperatures below and above the phase transition ($T_c$ = 535 K) for NaRu$_2$O$_4$. **(c)** The cell parameters with error bars of NaRu$_2$O$_4$ at different temperatures are extracted from single-crystal XRD data. The vertical line corresponds to $T_c$.

**Figure 3: The Converge Beam Electron Diffraction (CBED) patterns at room temperature in NaRu$_2$O$_4$ crystal**. **(a)** The room temperature CBED patterns for NaRu$_2$O$_4$ were taken along the reciprocal axis directions [100] (top panel) and **(b)** [130] (bottom panel). The black vertical solid lines are the position of the mirror plane (*m*), and solid red arrows are respective reciprocal axis directions. An arrowhead in (a, b) indicates the dynamical extinction of the glide plane and screw axis. **(c)** From the selected area electron diffraction pattern and CBED pattern along [001], we found whole pattern symmetry as 2*z*, which lacks both m$_x$ and m$_y$ symmetries and a c//2-fold symmetry. **(d)** Similarly, the diffraction pattern along [201] found the whole pattern symmetry without m$_y$.

**Figure 4: The HT and LT crystal structures**. **(a)** The HT-phase in NaRu$_2$O$_4$ has an orthorhombic structure with a *Pnma* space group, similar to the CaFe$_2$O$_4$-type structure [1]. The edge shared octahedral and Ru atoms are represented in light brown and O atoms in red color. **(b)** The HT-phase crystal structure (*Pnma*) of NaRu$_2$O$_4$ in *bc*-plane. **(c)** The LT-phase crystal structure of NaRu$_2$O$_4$ has monoclinic symmetry $P112_1/a$.

**Figure 5: Single-crystal x-ray diffraction (SC-XRD). (a)** Reciprocal lattice maps of the SC-XRD data for NaRu$_2$O$_4$ obtained at 470 K and 575 K with the first order transition at $T_c$ = 535 K. **(b)** The line cut for various temperatures below and above the first-order phase transition (> $T_c$) as extracted from the reciprocal space analysis. **(c)** Temperature dependence of the (4 3 0) super-lattice peak.

**Figure 6: Crystal structure of the tunnel compound NaRu$_2$O$_4$.** (a) Top view of the low-temperature structure. (b) Illustration of different types of zigzag ladders: "red" ladders with edge-sharing RuO$_6$ octahedra; "green" and "blue" ladders with corner-sharing octahedra. (c) The detailed structure of corner-sharing and edge-sharing ladders.

**Figure 7: Superstructure in the low-temperature phase and the results of the LSDA+*U* band structure calculations. (a)** The unit-cell of the low-temperature with two dimerized Ru bonds and the valence of Ru as determined from bond valence sum (BVS) calculations. **(b)** Structure of the blue corner-sharing ladder with different valence states of Ru and different bond lengths marked. **(c-f)** Orbital resolved DOS for four nonequivalent Ru atoms calculated within LSDA+*U* with *U* = 3.7 eV and *J* = 0.7 eV, assuming the charge order shown in (a). Ru orbitals are defined in local frames with the axes directed to the nearest O ions.

**Figure 8: Z-order in the two-leg corner- and edge-sharing RuO$_6$ zigzag ladders.** In the low-temperature phase, the sketch of self-organized Z-clusters formed within the edge-sharing red and corner-sharing blue ladders of NaRu$_2$O$_4$ lattice.

**Figure 9: CaFe$_2$O$_4$ type tunnel structure and edge shared RuO$_6$ 2 leg ladders in NaRu$_2$O$_4$ lattice. (a)** The orthorhombic *Pnma* crystal structure. **(b)** The edge-shared bond between the two RuO$_6$ octahedra for each dimer 1 (2.60 Å), dimer 2 (2.61 Å), and Z diagonal-bond (3.10 Å). **(c)** Schematic representation of the double chain with two-leg zigzag ladders (solid thick light-green, reddish-brown, and sky-blue lines). **(d)** The 2-leg zigzag ladders in two different forms, either corner (blue or green ladders) or edge (red ladders), shared RuO$_6$ octahedra.

**Table 1: The single-crystal x-ray diffraction measurement of NaRu$_2$O$_4$ at 300 K is used for the refinement.** The experimental structural result at room temperature has been determined by twin analysis (with twin (0.43748)) using the ShelX program [20]. The *P*112$_1$/*a* symmetry is used for the refinement as obtained from CBED Experiments. The reliable parameters of refinement are found to be: wR2 = 0.0821, R1 = 0.0344, GoF = S = 1.162: *a* = 99.256(1) Å, *b* = 5.634(1) Å, and *c* = 11.154(1) Å; γ = 90.10 (1)°; volume = 581.7(1) Å$^3$.

| Atoms | x | y | z | $U_{iso}$ | Site |
|---|---|---|---|---|---|
| Ru1(A) | 0.05988(6) | 0.10622(13) | 0.11842(5) | 0.0043(2) | 4e |
| Ru1(B) | 0.05614(7) | 0.64469(14) | 0.11228(4) | 0.0042(2) | 4e |
| Ru2(B) | 0.08283(7) | 0.14396(14) | 0.60006(5) | 0.0040(2) | 4e |
| Ru2(A) | 0.08712(7) | 0.60733(14) | 0.60517(4) | 0.0042(2) | 4e |
| Na1(A) | 0.2383(3) | 0.1310(9) | 0.3396(2) | 0.0160(6) | 4e |
| Na1(B) | 0.2412(3) | 0.6313(9) | 0.3417(3) | 0.0208(7) | 4e |
| O1A | 0.2874(5) | 0.1328(12) | 0.6631(4) | 0.0062(9) | 4e |
| O1B | 0.3030(5) | 0.6215(13) | 0.6541(4) | 0.0066(9) | 4e |
| O2A | 0.3870(5) | 0.1250(12) | 0.9847(4) | 0.0073(9) | 4e |
| O2B | 0.3829(5) | 0.6171(12) | 0.9676(4) | 0.0065(9) | 4e |
| O3A | 0.4676(5) | 0.1135(12) | 0.2149(4) | 0.0061(9) | 4e |
| O3B | 0.4783(5) | 0.6232(11) | 0.2240(4) | 0.0067(9) | 4e |
| O4A | 0.0951(5) | 0.1221(12) | 0.9385(4) | 0.0052(9) | 4e |
| O4B | 0.0775(5) | 0.6322(11) | 0.9310(4) | 0.0065(9) | 4e |

**Table 2: Expected coordinates at HT *Pnma* phase based on LT *P112$_1$/a* structure ( $a_0 \times b_0 \times c_0$). Atoms are at 4c site [19]**

| Atoms | x | 1/4 | z | -x+1/2 | 3/4 | z+1/2 | -x | 3/4 | -z | x+1/2 | 1/4 | -z+1/2 |
|---|---|---|---|---|---|---|---|---|---|---|---|---|
| Ru1 | 0.0580 | 0.25 | 0.1153 | 0.4420 | 0.75 | 0.6153 | -0.0580 | 0.75 | -0.1153 | 0.5580 | 0.25 | 0.3847 |
| Ru2 | 0.0850 | 0.25 | 0.6026 | 0.4150 | 0.75 | 0.1026 | -0.0850 | 0.75 | -0.6026 | 0.5850 | 0.25 | -0.1026 |
| Na1 | 0.2395 | 0.25 | 0.3409 | 0.2605 | 0.75 | -0.1591 | -0.2395 | 0.75 | -0.3409 | 0.7395 | 0.25 | 0.1591 |
| O1 | 0.2950 | 0.25 | 0.6585 | 0.2050 | 0.75 | 0.1585 | -0.2950 | 0.75 | -0.6585 | 0.7950 | 0.25 | -0.1585 |
| O2 | 0.3849 | 0.25 | 0.9762 | 0.1151 | 0.75 | 0.4762 | -0.3849 | 0.75 | -0.9762 | 0.8849 | 0.25 | -0.4762 |
| O3 | 0.4730 | 0.25 | 0.2195 | 0.0270 | 0.75 | -0.2805 | -0.4730 | 0.75 | -0.2195 | 0.9730 | 0.25 | 0.2805 |
| O4 | 0.0863 | 0.25 | 0.9347 | 0.4137 | 0.75 | 0.4347 | -0.0863 | 0.75 | -0.9347 | 0.5863 | 0.25 | -0.4347 |

**Table 3: Reported LT *Pnma* structure ($a_0 \times b_0 \times c_0$). Atoms are at 4c site [19]**

| Atoms | x | 1/4 | z | -x+1/2 | 3/4 | z+1/2 | -x | 3/4 | -z | x+1/2 | 1/4 | -z+1/2 |
|---|---|---|---|---|---|---|---|---|---|---|---|---|
| Ru1 | 0.0603 | 0.25 | 0.1152 | 0.4397 | 0.75 | 0.6152 | -0.0603 | 0.75 | -0.1152 | 0.5603 | 0.25 | 0.3848 |
| Ru2 | 0.0848 | 0.25 | 0.6036 | 0.4152 | 0.75 | 0.1036 | -0.0848 | 0.75 | -0.6036 | 0.5848 | 0.25 | -0.1036 |
| Na1 | 0.2399 | 0.25 | 0.3397 | 0.2601 | 0.75 | -0.1603 | -0.2399 | 0.75 | -0.3397 | 0.7399 | 0.25 | 0.1603 |
| O1 | 0.2946 | 0.25 | 0.6594 | 0.2054 | 0.75 | 0.1594 | -0.2946 | 0.75 | -0.6594 | 0.7946 | 0.25 | -0.1594 |
| O2 | 0.3847 | 0.25 | 0.9751 | 0.1153 | 0.75 | 0.4751 | -0.3847 | 0.75 | -0.9751 | 0.8847 | 0.25 | -0.4751 |
| O3 | 0.4730 | 0.25 | 0.2181 | 0.0270 | 0.75 | -0.2819 | -0.4730 | 0.75 | -0.2181 | 0.9730 | 0.25 | 0.2819 |
| O4 | 0.0870 | 0.25 | 0.9347 | 0.4130 | 0.75 | 0.4347 | -0.0870 | 0.75 | -0.9347 | 0.5870 | 0.25 | -0.4347 |

**Table 4: The four different models corresponding to four options for two dimers**: counting from the left, (3+, 4+) in the upper dimer, (3+, 4+) in the lower dimer. The summary of *Ru valences from the* Bond-Valence Sum (BVS) calculations for different Ru sites in NaRu$_2$O$_4$ at 300 K shows different models below. The model-3 looks consistent as per the observed dimerization on the two-leg zigzag ladder and our detailed LSDA+*U* calculations. Accordingly, we have used charges on the respective Ru sites in the NaRu$_2$O$_4$ crystal structure (Fig. 1 & 3a). The GII is the reliability parameter as Global Instability Index. The average valence on Ru sites is 3.5+.

| atoms | Model-1 | | | Model-2 | | |
|---|---|---|---|---|---|---|
| | assumed valence | result | GII | assumed valence | result | GII |
| Ru1(A) | 4+ | 3.5+ | 10.56% | 3+ | 2.9+ | 10.85% |
| Ru1(B) | 3+ | 3.1+ | | 4+ | 3.6+ | |
| Ru2(A) | 3+ | 2.9+ | | 4+ | 3.5+ | |
| Ru2(B) | 4+ | 3.6+ | | 3+ | 3.0+ | |
| atoms | Model-3 | | | Model-4 | | |
| | assumed valence | result | GII | assumed valence | result | GII |
| Ru1(A) | 3+ | 2.9+ | 10.43% | 4+ | 3.5+ | 11.02% |
| Ru1(B) | 4+ | 3.6+ | | 3+ | 3.1+ | |
| Ru2(A) | 3+ | 2.9+ | | 4+ | 3.5+ | |
| Ru2(B) | 4+ | 3.6+ | | 3+ | 3.0+ | |

Figure 1

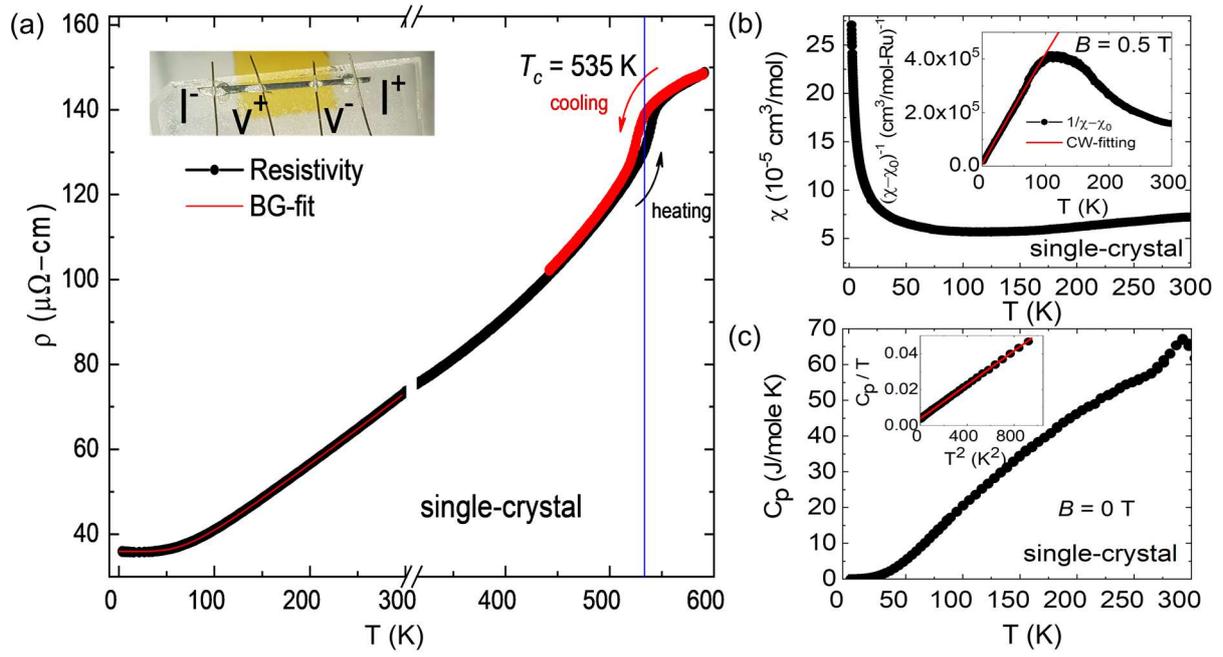

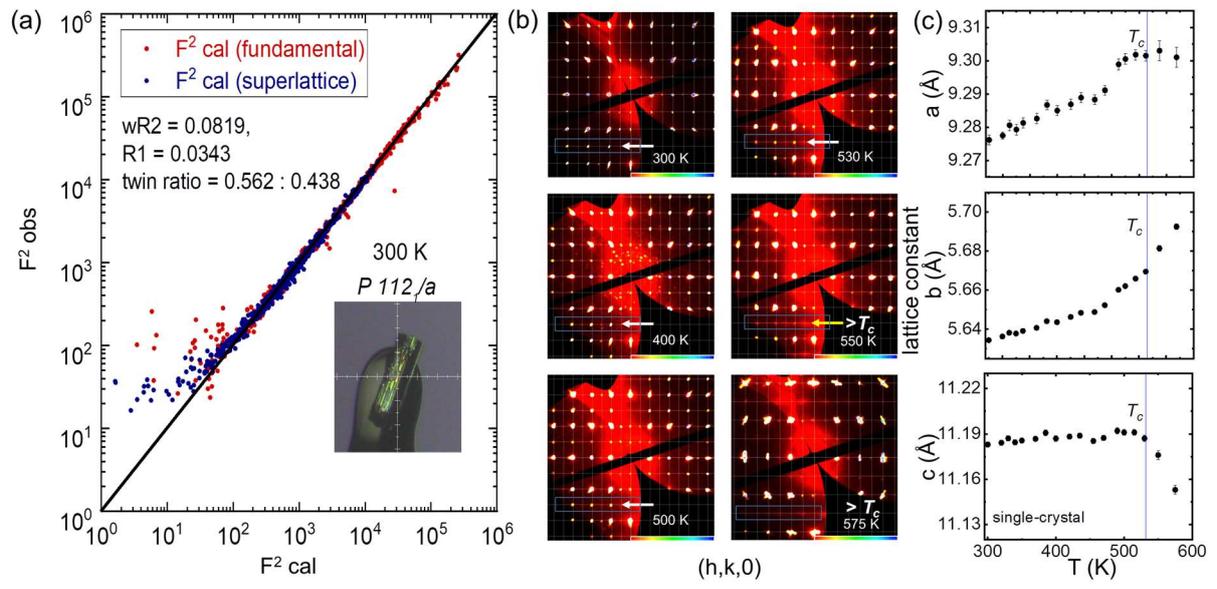

**Figure 3**

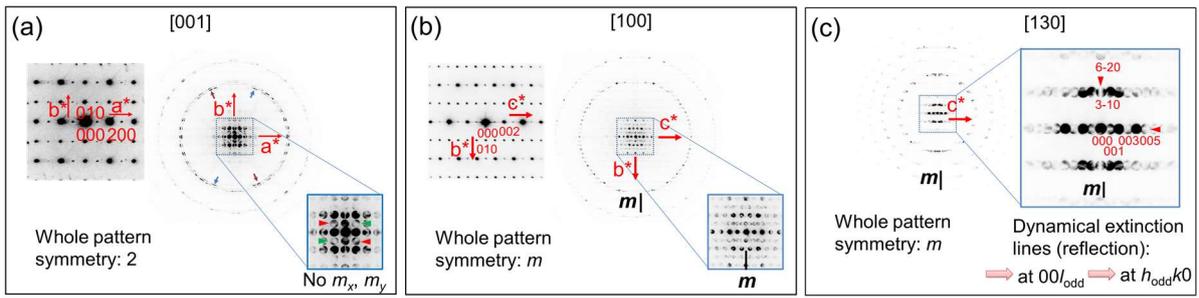

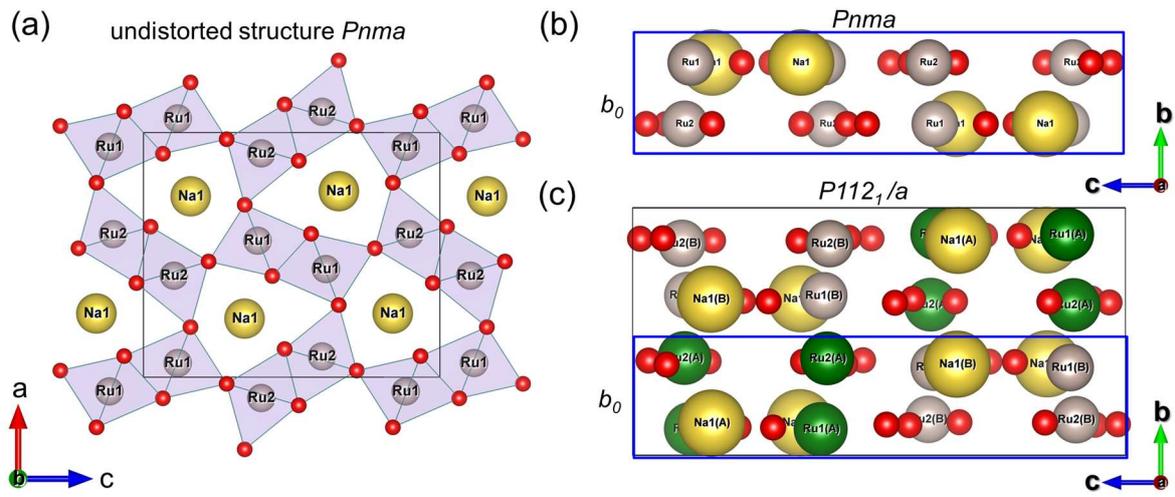

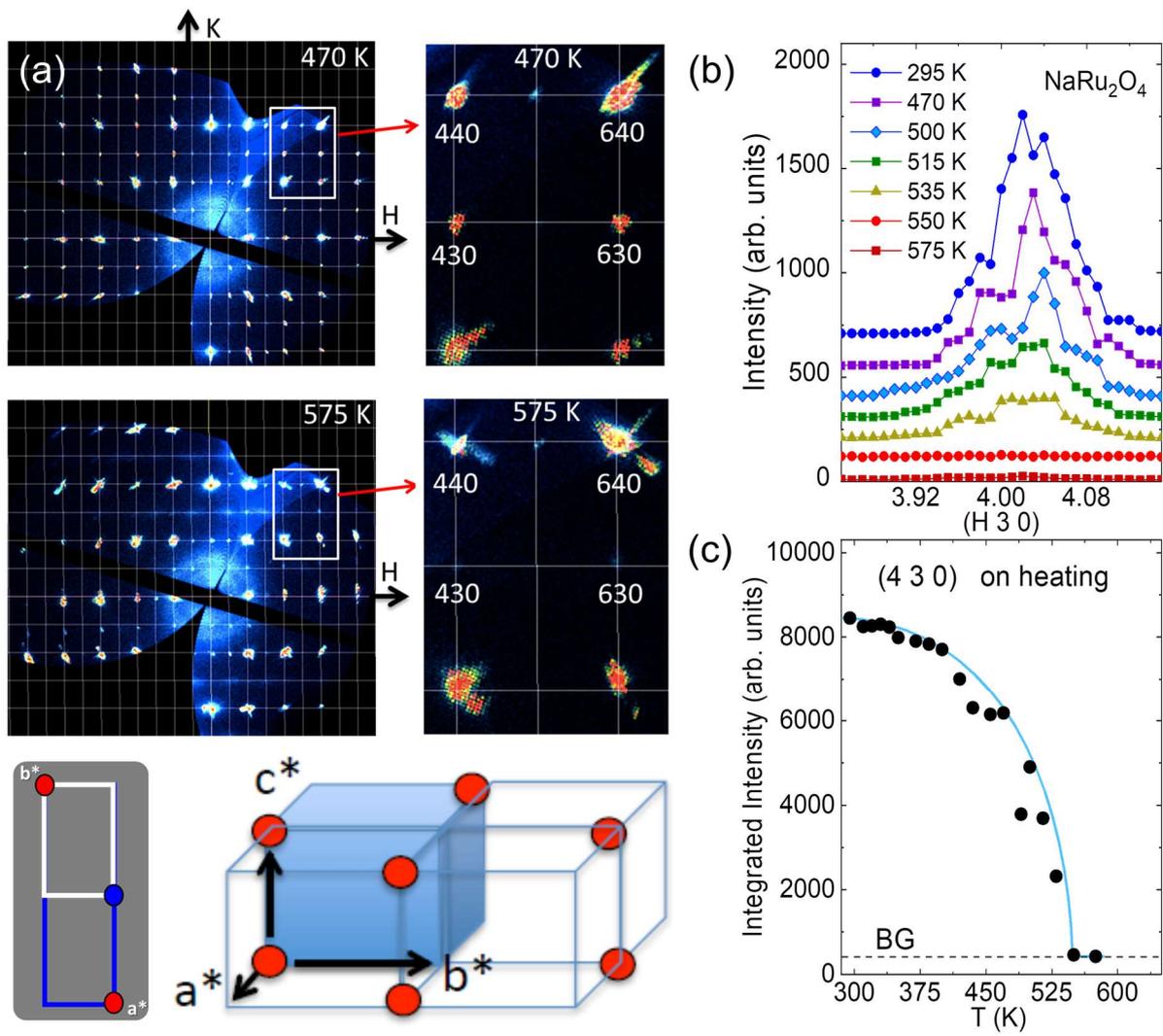

**Figure 6**

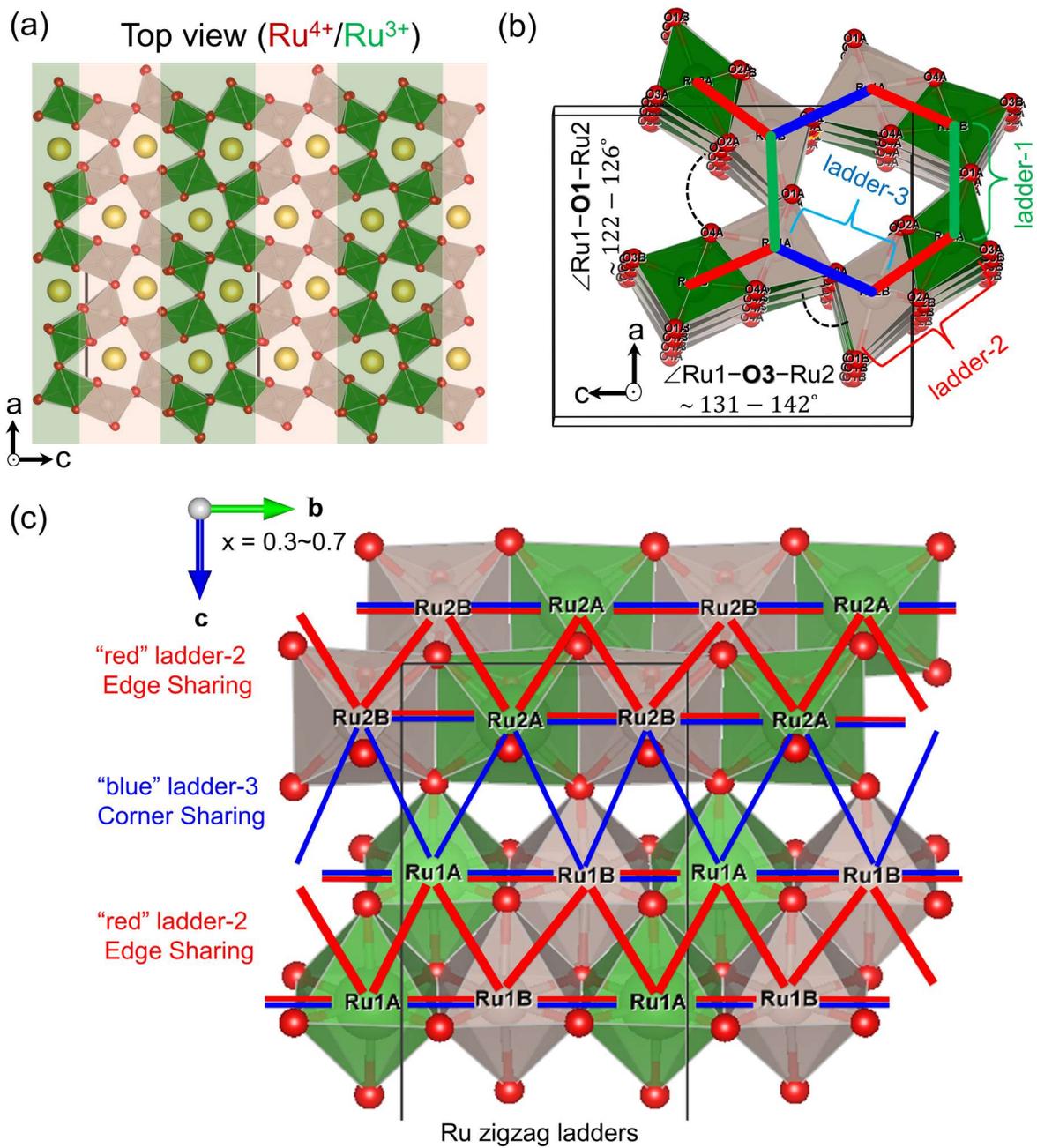

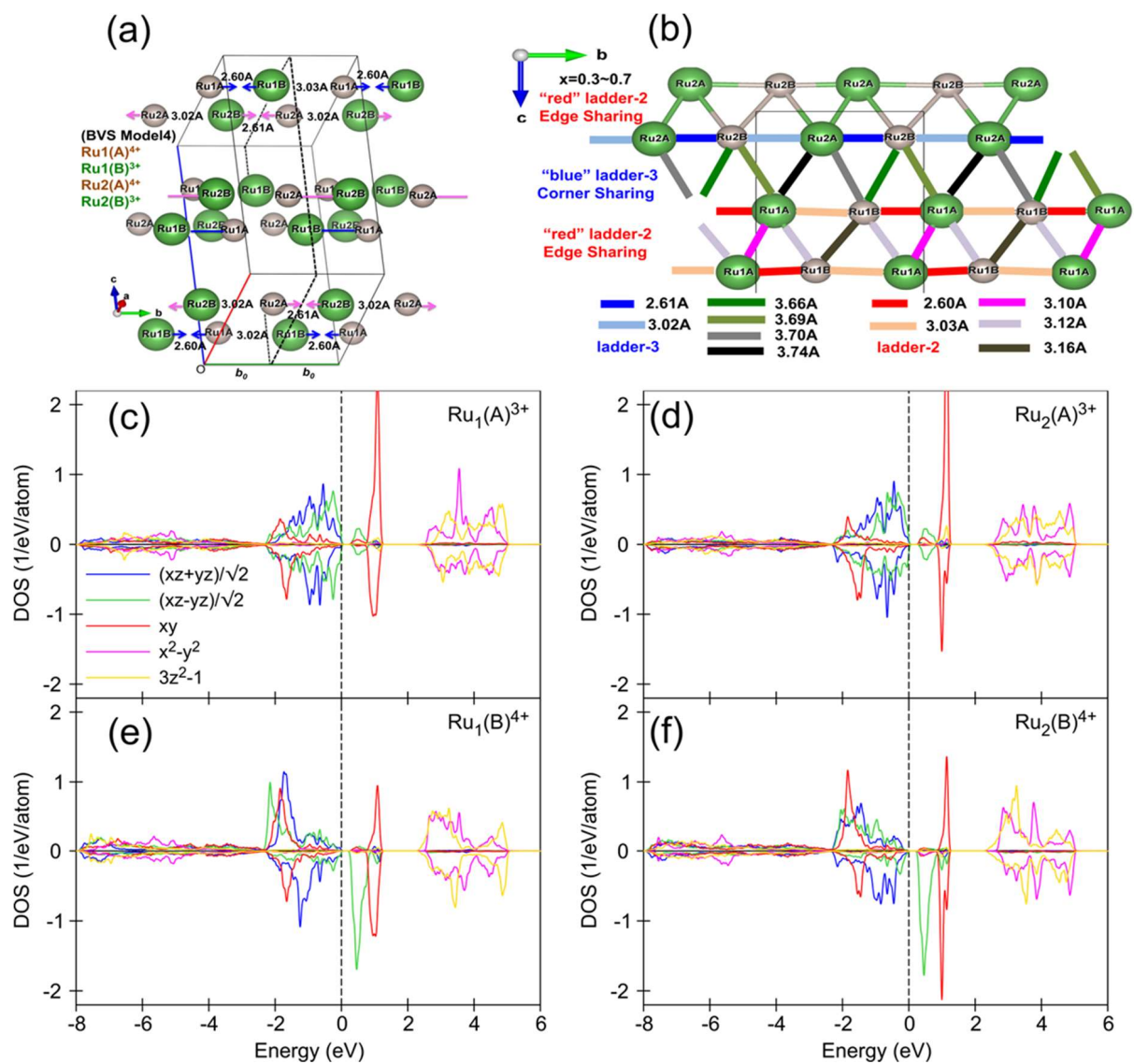

**Figure 8**

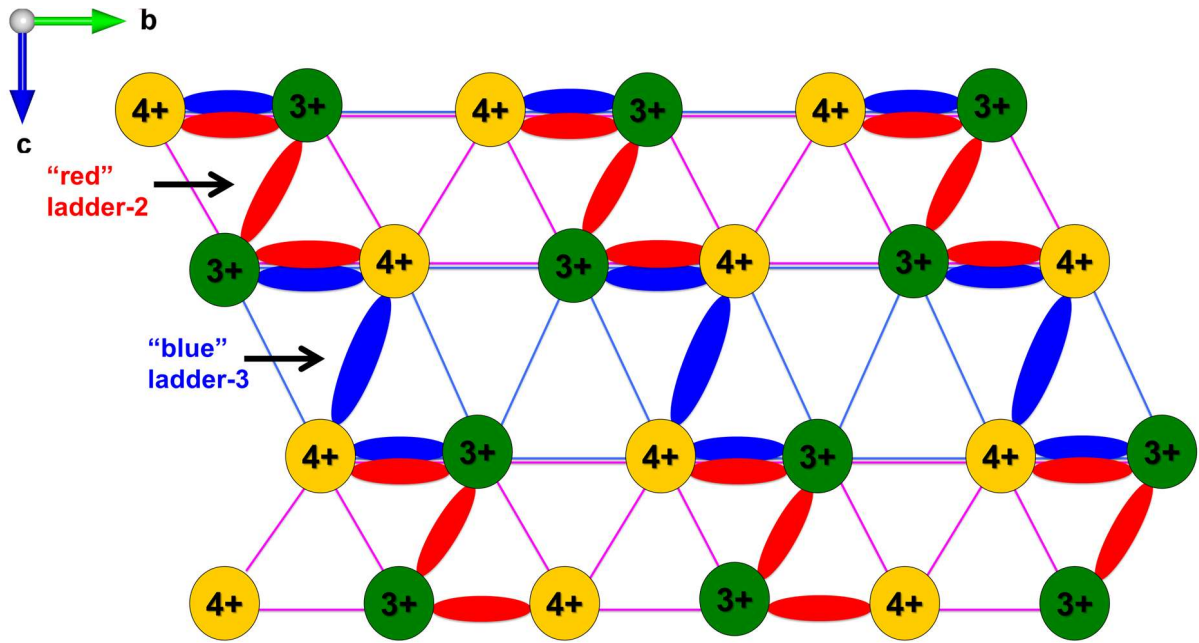

**Figure 9**

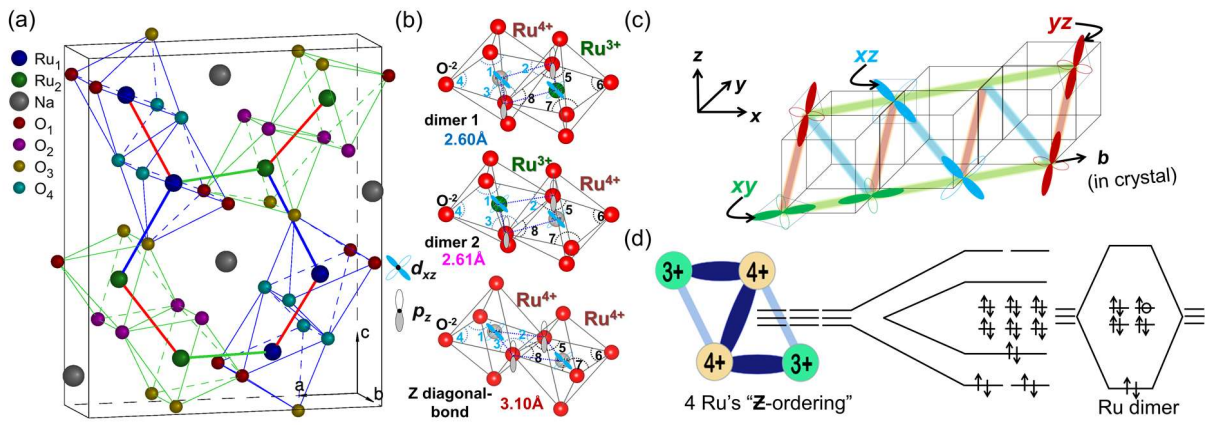

# Supplementary Information for

# Coexisting Z-type charge and bond order in metallic NaRu$_2$O$_4$

**Arvind Kumar Yogi**[1,2,3*,#], **Alexander Yaresko**[4*], **C. I. Sathish**[1,2], **Hasung Sim**[1,2], **Daisuke Morikawa**[5], **J. Nuss,**[4] **Kenji Tsuda**[6], **Y. Noda**[5,7], **Daniel I. Khomskii**[8#], **and Je-Geun Park**[1,2,9,10#]

[1]*Center for Correlated Electron Systems, Institute for Basic Science (IBS), Seoul 08826, Korea*
[2]*Department of Physics and Astronomy, Seoul National University, Seoul 08826, Korea*
[3]*UGC-DAE Consortium for Scientific Research, Indore-452001, India*
[4]*Max-Planck-Institut für Festkörperforschung, 70569 Stuttgart, Germany*
[5]*Institute of Multidisciplinary Research for Advanced Materials, Tohoku University, Sendai 980-8577, Japan*
[6]*Frontier Research Institute for Interdisciplinary Sciences, Tohoku University, Sendai 980-8578, Japan*
[7]*J-PARC Center, Japan Atomic Energy Agency, 2-4 Shirakata, Tokai, Ibaraki, 319-1195, Japan*
[8]*Institute of Physics II, University of Cologne, 50937 Cologne, Germany*
[9]*Center for Quantum Materials, Seoul National University, Seoul 08826, Korea*
[10]*Institute of Applied Physics, Seoul National University, Seoul 08826, Korea*

**I. Resistivity measurements**

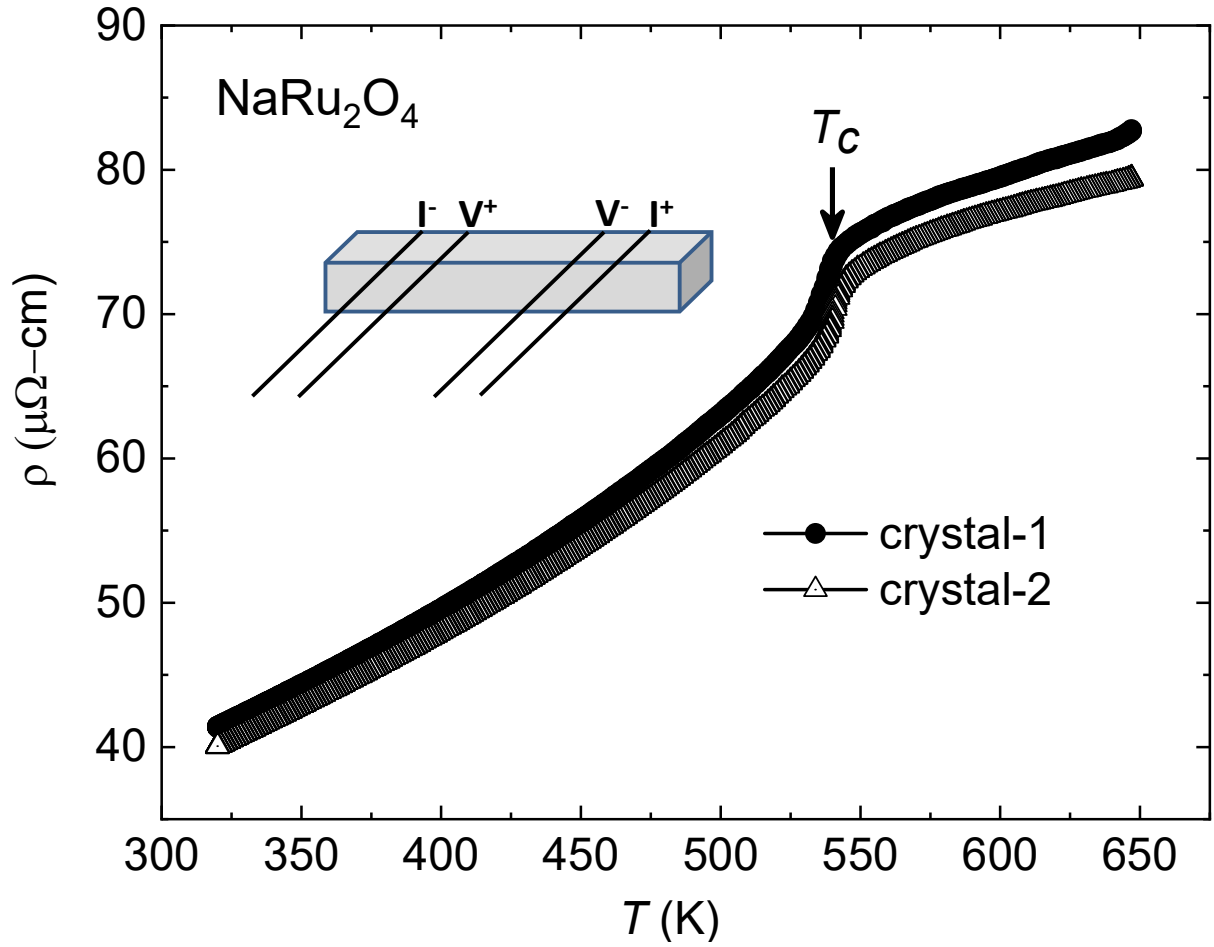

**Figure S1: Electrical transport measurements of NaRu$_2$O$_4$ on different single-crystals.** The temperature-dependent resistivity ($\rho$) curves are given as a function of temperature for two different single-crystals with a first-order phase transition at $T_C$ = 535 K, marked by a downward arrow.

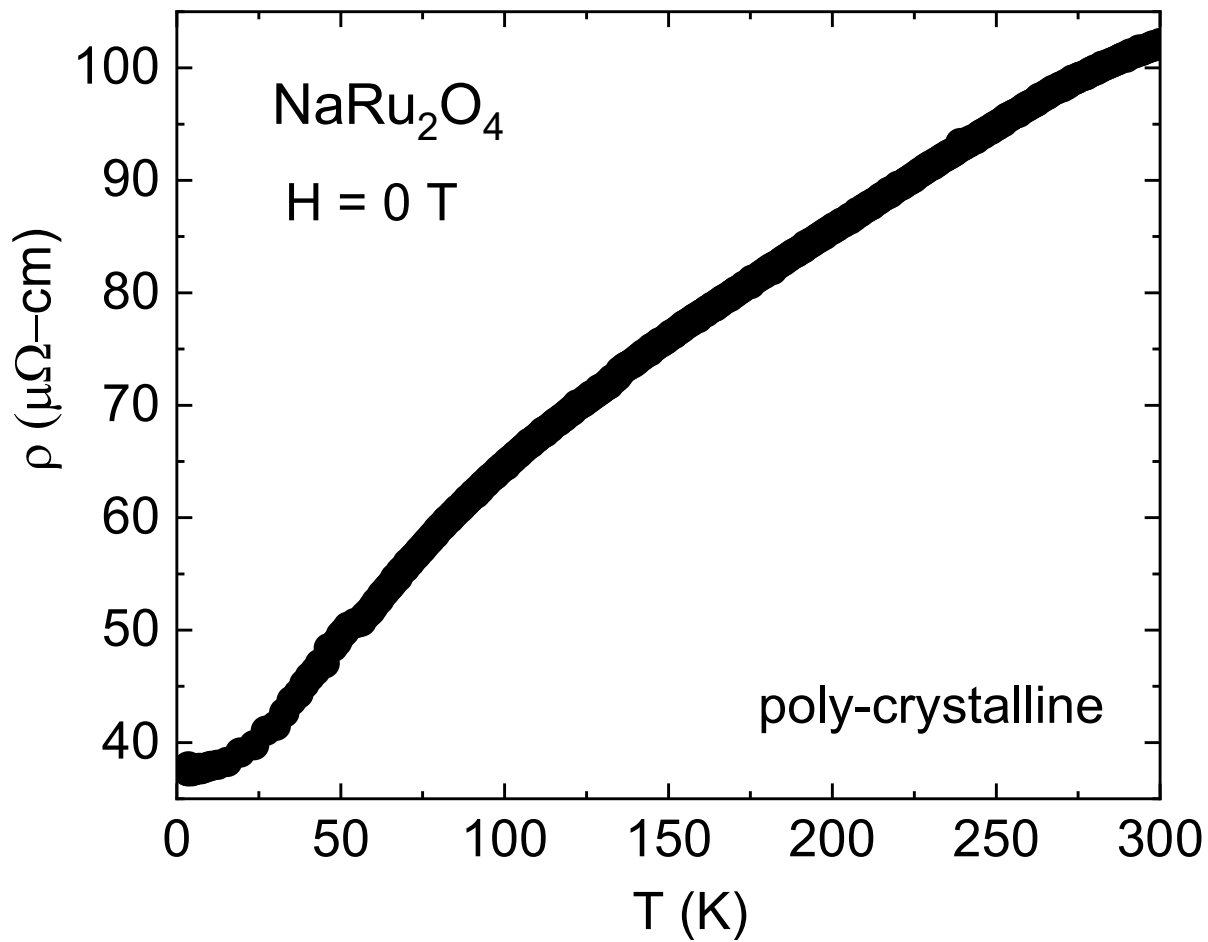

**Figure S2: Electrical transport measurement of NaRu$_2$O$_4$ on poly-crystals bar sample.** The temperature-dependent resistivity ($\rho$) curve is given as a function of temperature for poly-crystal rectunglar bar sample indicateing metallic nature in whole tempreture range.

**II. Modified Curie-Weiss fit**

We fitted the measured $\chi(T)$ by the following expression:

$$\chi(T) = \chi_0 + \frac{C}{(T + \theta_{CW})}$$

, where $\chi_0$ is the temperature-independent contribution.

The temperature dependence of the inverse susceptibility ($\chi^{-1}$) is strongly nonlinear, *i.e.*, it does not follow the Curie-Weiss behavior as shown in the inset of Fig. 1b. Regarding the Curie behavior at low temperature, we analyzed the data using a modified Curie-Weiss law in between 50 to 100 K. The fit yields the following values: Curie-Weiss temperature $\theta_{CW}$ = -1.8 K, observed moment $\mu_{eff}$ = 0.04 $\mu_B$ per Ru atoms and $\chi_0$ = 1.15×10$^{-7}$ (cm$^3$/ Ru mol)$^{-1}$, where the latter is the temperature-independent van Vleck contribution to the susceptibility. The CW-fit gives a significantly less observed moment, indicating negligible localized magnetic moments in NaRu$_2$O$_4$, similar to the reported polycrystalline sample [1].

**III. Merohedral twin refinement**

Our SC-XRD refinement using Fullprof software [2] reveals a bit large value of reliable parameters. Therefore, we turn to adapt the twin-law in our SC-XRD refinement using the Shelx software [3]. The twin-law defect is a symmetry operator of the crystal system. It is never the point group or Laue group of the crystal system. In the case of such existing twinning in the crystal system, one may get perfect overlap or close to the reflections from both domains, as shown by a schematics representation of reflections from twin domains in **Fig. S3**.

Moreover, the monoclinic lattice is crystallographically unique as it sometimes has $\beta$ angle very close to 90°. If twinning occurs, the unit cells in one domain may be rotated by 180° about the crystallographic *a*- or *c*-axes. Sometimes twinning does exist in a single-crystal due to energetically competitive inter-molecular interactions across twin domains. Due to this reason, twinning commonly occurs if a high-symmetry phase of material undergoes a phase transition to a lower-symmetry phase upon lowering or increasing of temperature [4]: a broken-symmetry element that is equivalent in the high-symmetry phase can show the twin-law in the low-symmetry phase. This is what we observed from our structural analysis. By applying the twin law [(-100), (010), and (00-1)], we improved the refinement results as well as SC-XRD data fitting. This means that the crystal measured was a merohedral twin, showing inversion symmetry in an experimentally determined space group (P112$_1$/a) in the lower Laue class. The observed twin ratio was found to be 0.43748 for NaRu$_2$O$_4$ crystal.

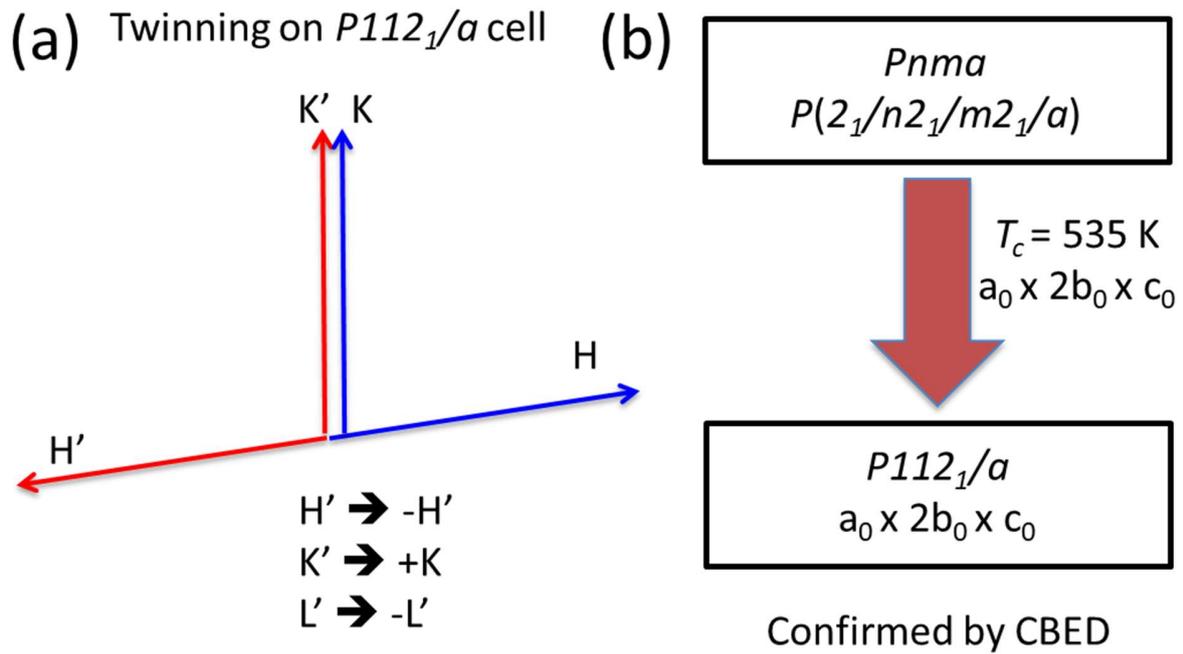

**Figure S3: The twin-law applied in NaRu$_2$O$_4$ crystal.** (a) The schematic representation of Bragg reflections from twin domains is shown by red and blue color arrows. Some possible twin reflections are demonstrated for H (H') and K (K') planes. (b) The phase transition ($T_c$ = 535 K) from HT-phase to LT-phase and cell doubling relation under group-subgroup is shown along the *b*-axis.

## IV. Frustrated 2-leg zigzag ladders in NaRu$_2$O$_4$

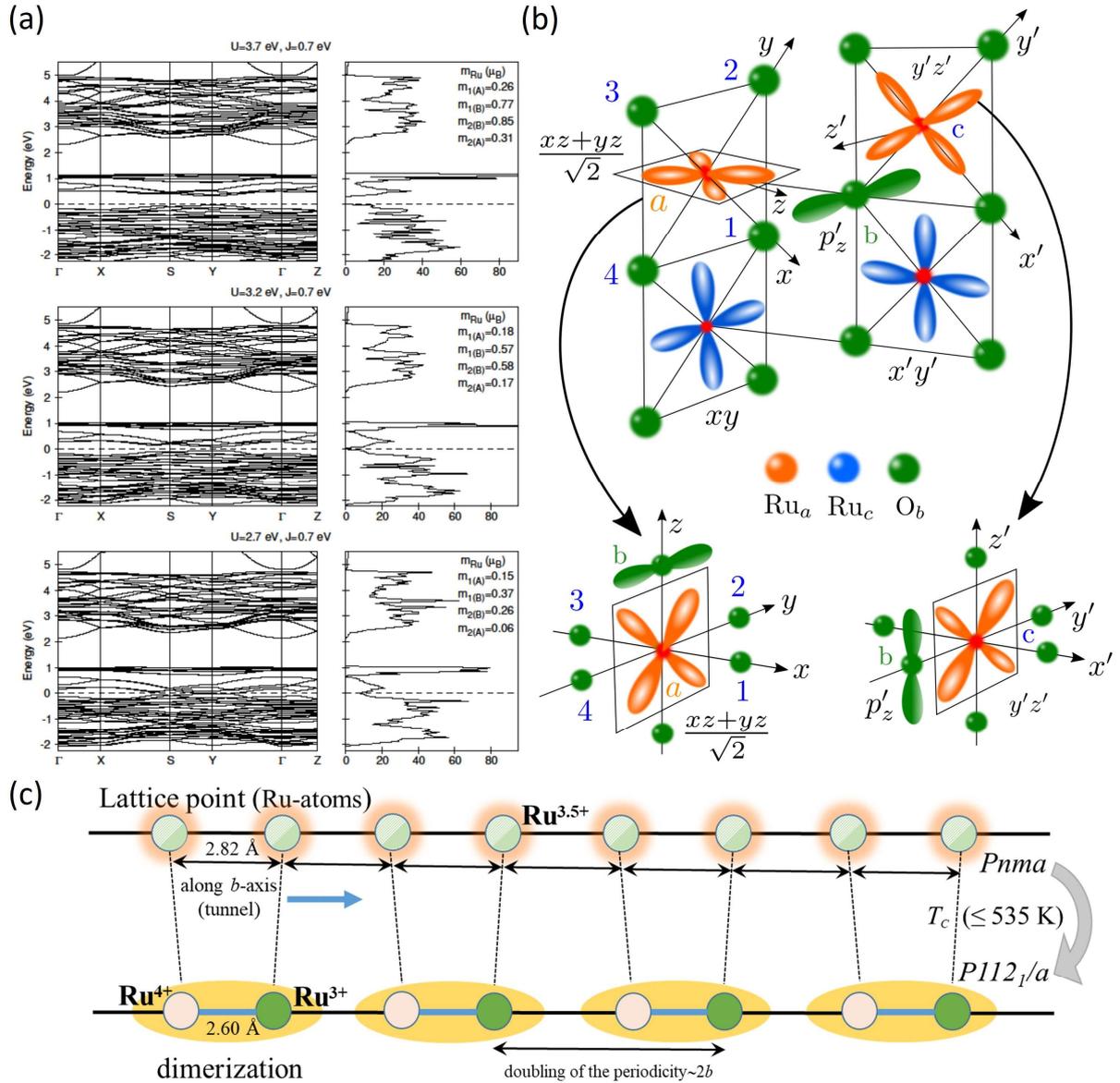

**Figure S4: Electronic structure, orbitals, and dimerization (Peierls-type) in 1D tunnel-lattice of NaRu$_2$O$_4$.** **(a)** NaRu$_2$O$_4$ band-structure is calculated by using LSDA+$U$, and the calculations were done for various sets of $U$ = 2.7, 3.2, and 3.7 and $J$ = 0.7 eV values ($U_{eff}$ = $U - J$ = 2.0, 2.5, and 3.0 eV). **(b)** The schematic representation of Ru orbitals in red and blue ladders. **(c)** The schematic of lattice distortion in NaRu$_2$O$_4$ below 535 K is shown in a metallic one-dimensional (1D) lattice.

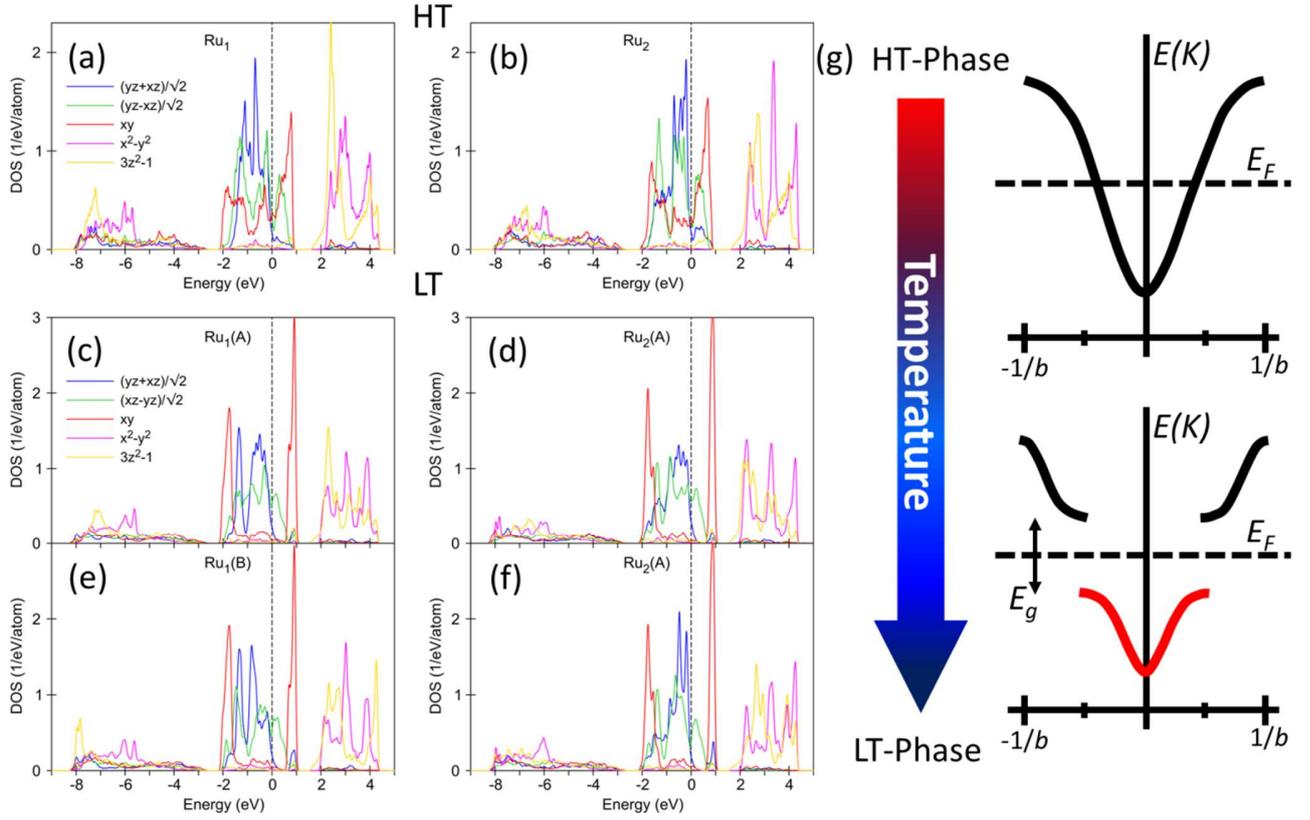

**Figure S5: Density of states (DOS) at high-temperature (HT) and low-temperature (LT) phase and consequence of Peierls transition in 1D tunnel-lattice of NaRu$_2$O$_4$. (a-f)** DOS for NaRu$_2$O$_4$ at HT- and LT-phase obtained from the LDA band-structure calculation. **(g)** The schematics of energy levels at HT-phase (top panel) and LT-phase (bottom panel) for NaRu$_2$O$_4$. The formation of energy-gap ($E_g$) at the Fermi level ($E_F$) due to unit-cell doubling along crystallographic *b*-axis breaks HT symmetry (*Pnma*) which results in Ruthenium (R) CO at LT-phase (*P112$_1$/a*).